\documentclass[preprint,showpacs,preprintnumbers,amsmath,amssymb]{revtex4}
\usepackage{graphicx}

\begin{document}

\title{Diffusion-controlled generation of a proton-motive force \\
across a biomembrane}

\author{ Anatoly Yu. Smirnov$^{1,2}$, Sergey E. Savel'ev$^{1,3}$, and Franco Nori$^{1,2}$ }

\affiliation{ $^1$ Advanced Science  Institute, The Institute of Physical and
Chemical Research (RIKEN), \\
Wako-shi, Saitama, 351-0198, Japan \\
$^2$ Center for Theoretical Physics, Physics Department, The University of Michigan, Ann Arbor, MI 48109-1040, USA \\
$^3$ Department of Physics, Loughborough University, Loughborough LE11 3TU, UK }

\date{\today}

\begin{abstract}
{ Respiration in bacteria involves a sequence of energetically-coupled electron and proton transfers creating an
 electrochemical gradient of protons (a proton-motive force) across the inner bacterial membrane. With a simple kinetic model
  we analyze a redox loop mechanism of proton-motive force generation mediated by a molecular shuttle diffusing inside the membrane.
This model, which includes six electron-binding and two proton-binding sites, reflects the main features of nitrate respiration in
\emph{E. coli} bacteria. We describe the time evolution of the proton translocation process. We find that the electron-proton
electrostatic coupling on the shuttle plays a significant role in the process of energy conversion between electron and proton
components. We determine the conditions where the redox loop mechanism is able to translocate protons against the transmembrane
voltage gradient above 200 mV  with a thermodynamic efficiency of about 37\%, in the physiologically important range of temperatures
from 250 to 350 K. }
\end{abstract}


 \maketitle

\section{Introduction}
Diffusion-controlled electron and proton transfer reactions are pivotal for the efficient energy transformation in respiratory
chains of animal cells and bacteria. During the process of respiration the energy, extracted from sunlight or from food molecules,
is converted into an electrochemical gradient of protons (also called a proton-motive force) across an inner mitochondrial or
bacterial membrane \cite{Alberts02,Skulachev88,GennisWikBook05,Nicholls92}. Thereafter, this energy is harnessed by ATP synthase for
a synthesis of adenosine triphosphate (ATP) molecules, the main energy currency of the cell. The energy stored in the proton
gradient can be also used to drive a rotation of a bacterial flagellar motor.

The energetically uphill translocation of protons is accomplished by a set of membrane-embedded proton pumps or by a redox loop mechanism
proposed in the original formulation of chemiosmotic theory \cite{Mitch76}. For a true proton pump (e.g., cytochrome c oxidase) electrogenic
events are associated with charges of protons crossing the membrane \cite{Skulachev88,GennisWikBook05}. In the redox loop mechanism the
transmembrane voltage is generated by electron charges moving across the membrane. This mechanism is responsible for a proton-motive force
generation in the respiratory chain of anaerobically grown bacteria such as the facultative anaerobe \emph{Escherichia coli}. In the absence of
oxygen and in the presence of nitrate, \emph{E. coli\ } can switch from oxidative respiration, which uses oxygen molecules as terminal electron
acceptors, to nitrate respiration, where nitrogen plays the role of a terminal acceptor of electrons in the process of nitrate-to-nitrite
reduction.

 The redox loop is formed by the formate dehydrogenase-N (\textit{Fdh-N}) enzyme and by the nitrate reductase enzyme (\textit{Nar})
(Fig.~1). The structures of these enzymes and positions of all redox centers have recently been determined
\cite{Jormakka02,Bertero03,RichSawers02,Jormakka03}. As a result of formate reduction, HCOO$^{-}$~$\rightarrow$~CO$_2$~+~H$^{+}$~+~2e$^{-}$, a
pair of high-energy electrons are delivered to the beginning of the pathway (source S) at the P-side of the inner (or plasma) membrane of
\emph{E.~coli}. Through the intermediate iron-sulfur clusters electrons are transferred, one after another, to the integral membrane subunit of
\textit{Fdh-N}, which includes hemes $b_P$ (site 1) and $b_C$ (site 2) located on the opposite sides of the membrane (see Fig.~1). The
subindices $P$ and $C$ here refer to ``Periplasm" and ``Cytoplasm", respectively.

\emph{E.~coli\ } utilizes a molecule of menaquinone (MQ) as a movable shuttle connecting the \textit{Fdh-N} and \textit{Nar} enzymes. Near the
N-side of the membrane menaquinone is populated with two electrons donated by heme $b_C$. In this process, menaquinone accepts two protons from
the N-side of the membrane turning into the form of menaquinol (MQH$_2$). The neutral menaquinol molecule diffuses to the P-side where it
donates two electrons to heme $b_L$ of the nitrate reductase and, simultaneously, two protons to the P-side proton reservoir.

Electrons are transferred, one by one, through heme $b_L$ (site 5), to heme $b_H$ (site 6) and, subsequently, through several iron-sulfur
clusters, to the site D on the cytoplasmic (N) space where the electrons reduce nitrate to nitrite,
NO$_3^{-}$~$\rightarrow$~NO$_2^{-}$~+~H$_2$O. The $L$ and $H$ subindices in the notations, $b_L$ and $b_H$, for the sites 5 and 6 refer to
``low" and ``high" redox potentials, respectively. Note that at the beginning of the electron transport chain (ETC), where formate is oxidized
to CO$_2$ and H$^{+}$, the midpoint redox potential is very low, $E_m = - 420$ mV. Thus, electrons entering ETC have high energies ($\sim 420$
meV). The menaquinone/menaquinol pair MQ/MQH$_2$ has a much higher redox potential, $E_m = - 80$ mV (and energy of order + 80 meV), which makes
possible the electron translocation against the transmembrane voltage. In the second half of the redox loop, formed by nitrate reductase,
electrons also move energetically downhill, from quinol ($E_m = - 80$ mV) to the nitrate reduction site having a midpoint potential, $E_m \sim +
420$ mV (and energy $\sim - 420$ meV) \cite{Blasco01}.

A geometrical disposition of the quinone-reducing center $b_C$ and the quinol-oxidizing center $b_L$ on opposite sites of the
membrane is crucial for the generation of the proton-motive force \cite{GennisWikBook05,Jormakka02,Bertero03}. Electrogenic events
resulting in net charge translocation occur when an electron moves from heme $b_P$ to heme $b_C$ in the \textit{Fdh-N} enzyme, and
from heme $b_L$ to heme $b_H$ located on the \textit{Nar} enzyme.

The crystal structures of the \textit{Fdh-N} and \textit{Nar} enzymes solved in Refs.\cite{Jormakka02,Bertero03} provide key components for
understanding the mechanism of proton-motive force generation through the redox loop. It should be emphasized, however, that the proton-motive
force generation is a dynamical process, so that structural analysis should be complemented by kinetic studies. For example, real time
investigations of electron and proton transfers in Complex I \cite{VerkhPNAS08} and Complex IV \cite{BelPNAS07} of mitochondria allow
elucidation of a time sequence of transfer events and get important information about electron and proton transition rates. Kinetic models of
the proton pumping processes in cytochrome $c$ oxidase \cite{Kim07,SugStuch08} and in bacteriorhodopsin \cite{Ferreira06} are also proven to be
beneficial for understanding experimental findings, as well as for an initiation of new experiments, giving a comprehensive picture of the
phenomenon.

In the present work we investigate a redox loop mechanism of a proton-motive force generation across the inner membrane of \emph{E.~coli\ }
bacterium within a simple physical model incorporating two hemes, $b_P$ and $b_C$, in the \textit{Fdh-N} enzyme, two hemes, $b_L$ and $b_H$, in
the \textit{Nar} enzyme, and a molecular shuttle (menaquinone) diffusing between these two halves of the redox loop. This diffusion is governed
by a Langevin equation. There is a pool of menaquinone/menaquinol molecules in the bacterial plasma membrane
\cite{Skulachev88,GennisWikBook05,Nicholls92}, but we only consider the contribution of a single menaquinone molecule to the electron and proton
translocation process. Because of this, the actual values of the electron and proton fluxes should be higher than the values calculated below.
In order to describe the process of loading/unloading the shuttle with electrons and protons, we employ a system of master equations, with
position-dependent transition rates between the shuttle and electron/proton reservoirs. With these equations we analyze the time dependence of
the proton-motive force generation process together with the dependence of numbers of transferred electrons and protons on a transmembrane
voltage and on temperature. A thermodynamic efficiency of the proton translocation across the inner bacterial membrane is defined and calculated
as well.

The paper is organized as follows. In Sec.~II we introduce a model of the system and present a set of master and Langevin equations, which
govern the time evolution of a proton translocation process. Sec.~III is devoted to a discussion of the key parameters of the model. In Sec.~IV
we report our main results and describe the steps for the kinetics of electron and proton transfer steps. The conclusions of the paper are
presented in Sec.~V.

\section{Model}

We take into consideration (see Fig.~1) six sites for an electron pathway through the system: two sites, 1 and 2, corresponding to hemes $b_P$
and $b_C$ of the \textit{Fdh-N} enzyme; two electron-binding sites, 3 and 4, on the menaquinone shuttle, and two sites, 5 and 6, related to
hemes $b_L$ and $b_H$ on nitrate reductase (\textit{Nar}). For the sake of simplicity we assume that heme $b_P$ (site 1), located on the
periplasmic (P) side of the membrane, is coupled to the source of electrons S, and that heme $b_H$ (site 6) having a high midpoint potential is
coupled to the electron drain D.

The source reservoir S characterized by an electrochemical potential $\mu_{\rm S}$ and the drain reservoir D described by an
electrochemical potential $\mu_{\rm D}$ provide a continuous flow of electrons through the electron transport chain (ETC). The
potential $\mu_{\rm S}$ roughly corresponds to the energy of electrons injected into the ETC after formate oxidation, $\mu_{\rm S}
\sim 420$ meV, whereas the drain potential $\mu_{\rm D}$ is related to the electron energy on the nitrate reduction site, $\mu_{\rm
D} \sim - 400$ meV. Note that we include the sign of the electron charge in the definition of the electron electrochemical
potential. This means that a site with a higher electron energy is characterized by a more negative redox midpoint potential
$E_{m}.$ Here, all energy parameters are measured in meV.

Taking into account two (instead of one) redox sites, 1 and 2, located on opposite sides of the membrane, allows us to describe the
process of transmembrane voltage generation during electron transfer (ET) along the \textit{Fdh-N} complex. Additional transmembrane
voltage is generated when an electron moves between two \textit{Nar} sites, 5 and 6, which are also located on the opposite sides of
the membrane.

The pathway for protons includes two proton-binding sites, 7 and 8, on the shuttle. We assume that the molecular shuttle moves along a line
connecting the redox sites 2 and 5. Depending on the position of the shuttle $x$ along this line, the proton-binding sites can be coupled either
to the positive or to the negative sides of the membrane (P- and N-proton reservoirs). The distributions of protons in the P and N reservoirs
are presumably described by the Fermi functions with the electrochemical potentials $\mu_{\rm P}$ (P-side) and $\mu_{\rm N}$ (N-side of the
membrane). In its completely reduced form of menaquinol MQH$_2$, the shuttle has a maximum load of two electrons and two protons, whereas in its
oxidized quinone form (denoted by MQ in Fig.~1) the shuttle is empty.

\subsection{Hamiltonian of the electron-proton system}
Within a formalism of secondary quantization \cite{Wingr93,Weinmann95,PREPump08,PREFlag08,PumpKinetics08} we introduce the creation
and annihilation Fermi operators, $a_{\alpha}^+, a_{\alpha}$, for an electron located on the site $\alpha \;(\alpha = 1,\ldots,6)$,
as well as the corresponding Fermi operators, $b_{\beta}^+, b_{\beta}$, for a proton on the protonable site $\beta \;(\beta = 7,8).$
The electron population of the $\alpha$--site is described by the operator $n_{\alpha} = a_{\alpha}^+a_{\alpha}$, whereas the proton
population of the $\beta$--site has the form: $n_{\beta} = b_{\beta}^+b_{\beta}$. Note that we use here methods of quantum transport
theory to derive classical master equations. A similar approach has been applied in studies of quantum coherence in biological
systems \cite{Guzik08}.

The main part of the system Hamiltonian, $H_0$, involves contributions from the energies, $\varepsilon_{\alpha}$, of electron sites
and energies, $\varepsilon_{\beta}$, of two proton-binding sites on the shuttle complemented by terms describing electrostatic
repulsions between sites 1 and 2 (with Coulomb energy $u_{12}$) and between sites 5 and 6 (with energy $u_{56})$. We also add  an
electron-electron Coulomb repulsion between two electron-binding sites, 3 and 4, on the shuttle (with an energy scale $u_{34}$) and
a term describing a repulsion between two protons, on the sites 7 and 8, occupying the shuttle (energy $u_{78}$). An electrostatic
attraction between electrons and protons travelling together on the menaquinol shuttle is described by the energy parameters
$u_{37},u_{38},u_{47},$ and $u_{48}$. As a result, the basic Hamiltonian $H_0$ of the electron-proton system has the form:
\begin{eqnarray}
H_0 &=& \sum_{\alpha=1}^6 \varepsilon_{\alpha} n_{\alpha} + \sum_{\beta=7}^8 \varepsilon_{\beta} n_{\beta} + u_{12}n_1 n_2 +
u_{34}n_3 n_4 + u_{56}n_5 n_6   \nonumber\\ &+& u_{78}n_7 n_8 - u_{37}n_3 n_7 - u_{38}n_3 n_8 - u_{47}n_4 n_7 - u_{48}n_4 n_8
\nonumber\\ &+& (n_3 + n_4 - n_7 - n_8)^2 U_s(x).\label{H0}
\end{eqnarray}
The last term in Eq.~(\ref{H0}), which depends on the shuttle position $x$,  describes the contribution of a potential barrier
$U_s(x)$, which prevents a charged shuttle from crossing the interior of the lipid membrane. The barrier has an almost rectangular
shape,
\begin{equation}
U_s(x) = U_{s0} \left\{\left[\exp\left( \frac{x - x_s}{l_s} \right) + 1 \right]^{-1} -   \left[ \exp\left( \frac{x + x_s}{l_s} \right) +
1\right]^{-1} \right\}, \label{Us}
\end{equation}
with a height $U_{s0}$, a steepness $l_s$, and a width $2x_s$. This is multiplied by the shuttle charge squared: $(n_3 + n_4 - n_7 -
n_8)^2$. The height $U_{s0}$ of this potential is roughly equal to the energy penalty (in meV) for moving a molecule with a charge
$q_0$ (in units of $|e|$) and a radius $r_0$ (in nm) from a medium with a dielectric constant $\epsilon_1$  to a medium with a
constant $\epsilon_2$ \cite{Armen06},
\begin{equation}
U_{s0} = \frac{1440 \, q_0^2}{2 r_0} \left(\frac{1}{\epsilon_2} - \frac{1}{\epsilon_1}\right).
\end{equation}
For example, the transfer of a charged molecule ($q_0=1$) with  radius $r_0 = 0.3$ nm, from water ($\epsilon_1=80$) to the lipid
membrane with $\epsilon_2=3$, results in the dielectric penalty $U_{s0}~=~770$ meV. The specific shape of the barrier $U_s(x)$ in
Eq.~(\ref{Us}) is of little importance for the results from this model.

Electrons in the source (drain) reservoir are described by the creation and annihilation operators $c_{k{\rm S}}^{+}, c_{k{\rm S}}$
($c_{k{\rm D}}^{+}, c_{k{\rm D}}$), and for protons in the N (P) reservoir we introduce operators $d_{q{\rm N}}^{+}, d_{q{\rm N}}$
($d_{q{\rm P}}^{+}, d_{q{\rm P}}$), so that the Hamiltonian of the electron source and drain reservoirs, $H_{\rm SD}$, and the
Hamiltonian of the proton reservoirs, $H_{\rm NP}$, can be expressed as
\begin{eqnarray}
H_{\rm SD} &=& \sum_k ( \varepsilon_{k{\rm S}} c_{k{\rm S}}^{+}c_{k{\rm S}} + \varepsilon_{k{\rm D}} c_{k{\rm D}}^{+}c_{k{\rm D}}), \nonumber\\
H_{\rm NP} &=& \sum_q ( \varepsilon_{q{\rm N}} d_{q{\rm N}}^{+}d_{q{\rm N}} + \varepsilon_{q{\rm P}} d_{q{\rm P}}^{+}d_{q{\rm P}}).
\label{Hleads}
\end{eqnarray}
Here, $\varepsilon_{k{\rm S}}$ and $\varepsilon_{k{\rm D}}$ are the energies of the electrons in the S and D reservoirs, and depend
on the quasi-momentum parameter $k$. The energies of the protons in the N- and P-reservoirs, $\varepsilon_{q{\rm N}}$ and
$\varepsilon_{q{\rm P}}$, depend on another continuous parameter $q$.

Electrons in the source and drain reservoirs ($\varsigma$ = S,D) and protons on the negative and positive ($\sigma$ = N,P) sides of the membrane
can be characterized by the corresponding Fermi distributions, $f_{\varsigma} (\varepsilon_{k \varsigma }) $ and $ F_{\sigma} (\varepsilon_{q
\sigma})$:
\begin{eqnarray}
f_{\varsigma}(\varepsilon_{k \varsigma}) &=& \left[\exp\left( \frac{\varepsilon_{k \varsigma }-\mu_{\varsigma}}{T}\right) + 1\right]^{-1}, \nonumber\\
F_{\sigma}(\varepsilon_{q \sigma}) &=& \left[\exp\left( \frac{\varepsilon_{q \sigma}-\mu_{\sigma}}{T}\right) + 1\right]^{-1}. \label{Fermi}
\end{eqnarray}
 We introduce here the electrochemical potentials $\mu_{\sigma }$  of the proton reservoirs and the potentials $\mu_{\varsigma}$ for the
electron source and drain. The potential $\mu_{\rm S}$ is related to the highest occupied energy level of the molecular complex S
supplying the ETC with electrons, and the potential $\mu_{\rm D}$  plays a similar role for the molecular complex D providing an
electron outflow.

Couplings between the electron site 1 (heme $b_P$) and the source S, and between the site 6 (heme $b_H$) and the electron drain D are determined
by the Hamiltonian
\begin{equation}
H_{\rm e} =  - \sum t_{k{\rm S}}\, c_{k{\rm S}}^+\,a_1  -  \sum t_{k{\rm D}}\, c_{k{\rm D}}^+\, a_6 + H.c., \label{He}
\end{equation}
with the corresponding transition coefficients $t_{k{\rm S}}$ and $t_{k{\rm D}}$. The similar Hamiltonian describes proton transitions between
the shuttle and the proton reservoirs,
\begin{equation}
H_{\rm p} =  - \sum (T_{q{\rm N}}\, d_{q{\rm N}}^+ + T_{q{\rm P}} \, d_{q{\rm P}}^+)(b_7 + b_8)   + H.c. \label{Hp}
\end{equation}
Here, the coefficients $T_{q{\rm N}}$ and $T_{q{\rm P}}$, which are assumed to be the same for both sites 7 and 8, depend on the shuttle
position $x$. The transitions between the redox sites 1, 6 and the electron source S and drain D as well as between the N- and P-sides of the
membrane and the protonable sites 7, 8 on the shuttle are determined by the energy-independent electron and proton rates
\cite{Wingr93,Weinmann95,PREPump08,PREFlag08,PumpKinetics08}
\begin{eqnarray}
\gamma_{\varsigma} = 2\pi \sum_k |t_{k \varsigma}|^2 \delta (E - \varepsilon_{k\varsigma} ), \nonumber\\
 \Gamma_{\sigma} = 2\pi \sum_q |T_{q\sigma}|^2
\delta (E - E_{q\sigma} ). \label{gammaSDNP}
\end{eqnarray}
The proton transition rates $\Gamma_{\rm N},\, \Gamma_{\rm P}$ depend on the distances (either $x+x_0$ or  $x_0 -x$) between the
shuttle and the N or P-sides of the membrane:
\begin{eqnarray}
\Gamma_{\rm N} &=& \Gamma_{{\rm N}0} \left[ \exp\left( \frac{x+x_0}{l_p}\right) + 1 \right]^{-2}, \nonumber\\
\Gamma_{\rm P} &=& \Gamma_{{\rm P}0} \left[ \exp\left( \frac{x_0-x}{l_p}\right) + 1 \right]^{-2}, \label{GammaX}
\end{eqnarray}
where $x = x(t)$ is the coordinate of the shuttle and $l_p$ is the proton transition length.

The electron tunneling between the redox centers $1,\ldots,6$ is governed by the Hamiltonian $H_{\rm tun}$,
\begin{eqnarray}
H_{\rm tun} &=& - \Delta_{12} a_1^{+}a_2 - \Delta_{23} a_2^{+}a_3 - \Delta_{24} a_2^{+}a_4   \nonumber\\ &-& \Delta_{35} a_3^{+}a_5 -
\Delta_{45} a_4^{+}a_5 - \Delta_{56} a_5^{+}a_6 + H.c. \label{Htun}
\end{eqnarray}
The electrons are transferred between the site 2, located at $x = -x_0$, and the electron-binding sites 3 and 4 on the shuttle. On the opposite
side of the membrane, at $x = x_0$, the electrons tunnel from the sites 3 and 4 to the site 5. These transfers drastically depend on the shuttle
position $x$. According to quantum mechanics, we can model the position dependence of the tunneling coefficients by the exponential functions:
\begin{eqnarray}
|\Delta_{23}|^2 = |\Delta_{24}|^2 = |\Delta_{2}|^2 \exp\left(- 2\ \frac{|x+x_0|}{l_e}\right), \nonumber\\
|\Delta_{35}|^2 = |\Delta_{45}|^2 = |\Delta_{5}|^2 \exp\left(- 2\ \frac{|x-x_0|}{l_e}\right), \label{DeltaX}
\end{eqnarray}
where $l_e$ is an electron tunneling length.

\subsection{Environment}

The atomic motion of the protein medium has a significant effect on electron charge transfer between the active sites. Usually (see
Refs.~\cite{Garg85,Krish01,Weiss08}) the environment is represented as a collection of independent harmonic oscillators. The coupling of these
oscillators to electronic degrees of freedom can be described by the Hamiltonian $H_{\rm env}$,
\begin{eqnarray}
H_{\rm env} = \sum_j \frac{p_j^2}{2m_j} + \frac{1}{2} \sum_j m_j\omega_j^2 \left(x_j - \sum_{\alpha=1}^6 x_{j\alpha}n_{\alpha} - x_{j{\rm S}}
n_{\rm S} - x_{j{\rm D}} n_{\rm D}  \right)^2. \label{Henv}
\end{eqnarray}
Here, $x_j$ and $p_j$ are the position and momentum of the $j$-oscillator, having mass $m_j$ and a frequency $\omega_j$. Also,
$n_{\rm S} = \sum_k c_{k{\rm S}}^{+}c_{k{\rm S}}$ and $n_{\rm D} = \sum_k c_{k{\rm D}}^{+}c_{k{\rm D}}$ are the total populations of
the source and drain reservoirs; $x_{j\alpha}, x_{j{\rm S}}, x_{j{\rm D}}$ are the set of coupling constants between electrons and
their surroundings.

Thus, the total Hamiltonian of the system has the form
\begin{equation}
H = H_0 + H_{\rm SD} + H_{\rm NP} + H_{\rm e}  + H_{\rm p} + H_{\rm tun} + H_{\rm env}. \label{Htot}
\end{equation}
A unitary transformation, $H' = U^{+}HU,$ with
\begin{equation}
U = \exp\left\{ - i \sum_j p_j \left( \sum_{\alpha} x_{j\alpha}n_{\alpha} +  x_{j{\rm S}} n_{\rm S} + x_{j{\rm D}} n_{\rm D}  \right) \right\},
\label{UT}
\end{equation}
applied to the Hamiltonian $H$, removes the environment variables \{$x_j$\} from the Hamiltonian $H_{\rm env}$ and introduces phase
shifts into the tunneling Hamiltonian $H_{\rm tun}$:
\begin{eqnarray}
H' &=& H_0 + H_{\rm SD} + H_{\rm NP} + H_{\rm e}  + H_{\rm p} + H_{\rm tun}' + \sum_j \left( \frac{p_j^2}{2m_j} + \frac{m_j\omega_j^2
x_j^2}{2}\right), \label{Hnew}
\end{eqnarray}
where
\begin{eqnarray}
H_{\rm tun}' = - Q_{12} a_1^{+}a_2 - Q_{23} a_2^{+}a_3 - Q_{24} a_2^{+}a_4  - \nonumber\\ Q_{35} a_3^{+}a_5 - Q_{45} a_4^{+}a_5 - Q_{56}
a_5^{+}a_6 + H.c., \label{HtunNew}
\end{eqnarray}
is a new tunneling Hamiltonian, and
\begin{eqnarray}
Q_{\alpha\alpha'} = Q_{\alpha'\alpha}^+ = \Delta_{\alpha \alpha'} \exp\{ i \sum_j p_j(x_{j\alpha} - x_{j\alpha'})\},  \label{QQ}
\end{eqnarray}
is a phase shift corresponding to the electron transition from site $\alpha'$ to site $\alpha$ \,($\hbar = 1$). For simplicity, we neglect here
the phase shifts for transitions between the source reservoir and the site 1, $x_{j{\rm S}} = x_{j 1}$, and between the site 6 and the electron
drain, $ x_{j 6} = x_{j{\rm D}}$, together with shifts related to proton transfers. The electron and proton reservoirs are described by
continuous energy spectra. The broadening of the reservoir energy states allows non-resonant transitions, e.g., between site 1 and the source S,
thus reproducing some effects of the corresponding (1-to-S) phase shifts. Recall also that the tunneling rates $\Delta_{\alpha\alpha'}$ for
transitions between the sites 2 and 3, 2 and 4, 3 and 5, 4 and 5 depend on the shuttle position $x(t)$ and, thus, depend on time (see
Eq.~(\ref{DeltaX})). However, this time dependence is much slower than the time variations of environment-induced phase factors.

\subsection{Basis states}
To describe all possible occupational configurations of the electron-proton system, we introduce a basis of 256 eigenstates,
$|\mu\rangle,$ of the Hamiltonian $H_0$: $H_0 |\mu\rangle = E_{\mu} |\mu\rangle,\, \mu = 1,\ldots,256,$ characterized by the energy
spectrum $E_{\mu}.$  The basis begins with the vacuum state, where there are no particles on the sites $1,\ldots,8$: $|1\rangle =
|0_1 0_2 0_3 0_4 0_5 0_6 0_7 0_8\rangle$, and finally ends with the state $|256\rangle$ describing the fully-populated system:
$|256\rangle = |1_1 1_2 1_3 1_4 1_5 1_6 1_7 1_8\rangle$. Here, the notation $0_{\alpha} (1_{\alpha})$ means that the electron site
$\alpha$ is empty (occupied). Similar notations are introduced for the proton sites 7 and 8.

It is of interest that all operators of the system, except the operators of the electron and proton reservoirs, can be expressed in terms of the
basic Heisenberg operators $\rho_{\mu\nu} = |\mu\rangle\langle \nu|$, for example,
\begin{eqnarray}
a_{\alpha}^+a_{\alpha'} = \sum_{\mu\nu} (a_{\alpha}^+a_{\alpha'})_{\mu\nu} \rho_{\mu\nu}, \nonumber\\
 a_{\alpha} = \sum_{\mu\nu} a_{\alpha;\mu\nu} \rho_{\mu\nu}, \,\, b_{\beta} = \sum_{\mu\nu} b_{\beta;\mu\nu} \rho_{\mu\nu}, \label{abRho}
\end{eqnarray}
where $\alpha , \alpha' = 1,\ldots,6;$ $ \beta = 7,8;$ and $$a_{\alpha;\mu\nu} = \langle \mu |a_{\alpha}|\nu\rangle,\;
b_{\beta;\mu\nu} = \langle \mu |b_{\beta}|\nu\rangle$$ are the matrix elements of the electron and proton operators in the basis
$|\mu\rangle.$
 The Hamiltonian $H_0$ has a diagonal form,
\begin{equation}
H_0 = \sum_{\mu =1}^{256} E_{\mu} \rho_{\mu}, \label{H0Mu}
\end{equation}
whereas the tunneling Hamiltonian $H_{\rm tun}$ (we drop hereafter a prime sign) has only off-diagonal elements,
\begin{equation}
H_{\rm tun} = - \sum_{\mu\nu} {\cal A}_{\mu\nu} \rho_{\mu\nu} + H.c.. \label{HtunMuNu}
\end{equation}
Here $\rho_{\mu}$ denotes a diagonal operator, $\rho_{\mu} \equiv \rho_{\mu\mu} = |\mu\rangle\langle \mu|$, and ${\cal A}_{\mu\nu}$
is a combination of operators, describing the environment,
\begin{eqnarray}
{\cal A}_{\mu\nu} &=& Q_{12} (a_1^{+}a_2)_{\mu\nu} + Q_{23} (a_2^{+}a_3)_{\mu\nu} + Q_{24} (a_2^{+}a_4)_{\mu\nu}  \nonumber\\ &+& Q_{35}
(a_3^{+}a_5)_{\mu\nu} + Q_{45} (a_4^{+}a_5)_{\mu\nu} + Q_{56} (a_5^{+}a_6)_{\mu\nu}. \label{Amunu}
\end{eqnarray}
The Hamiltonian $H_{\rm e}$, modelling the electron transfer from the source and drain to the sites 1 and 6, and the Hamiltonian
$H_{\rm p}$, which is responsible for proton transitions between the shuttle and the proton reservoirs, are also expressed in terms
of the basis matrix $\rho_{\mu\nu}$,
\begin{eqnarray}
H_{\rm e} &=&  - \sum_k \sum_{\mu\nu}  ( t_{k{\rm S}}  c_{k{\rm S}}^+a_{1;\mu\nu}  +   t_{k{\rm D}} c_{k{\rm D}}^+a_{6;\mu\nu}) \rho_{\mu\nu} +
H.c. \nonumber\\
H_{\rm p} &=&  - \sum_q  \sum_{\mu\nu} ( T_{q{\rm N}} d_{q{\rm N}}^+ + T_{q{\rm P}} d_{q{\rm P}}^+)(b_{7;\mu\nu} + b_{8;\mu\nu}) \rho_{\mu\nu} +
H.c. \label{HepMuNu}
\end{eqnarray}

\subsection{Master equation}

The average value, $\langle \rho_{\mu}\rangle$,  of the operator $\rho_{\mu}$ determines the probability to find the system in the
state $|\mu\rangle$. This probability can be found from the Heisenberg equation,
\begin{equation}
\dot{\rho}_{\mu} = -i [\rho_{\mu},H_{\rm e} + H_{\rm p}]_{-} - i [\rho_{\mu}, H_{\rm tun}]_{-}, \label{HeisEq}
\end{equation}
averaged over the states of reservoirs and over fluctuations of the environment. It is convenient to employ methods of quantum
transport theory and the theory of open quantum systems \cite{Wingr93,Weinmann95, PREPump08,PREFlag08,PumpKinetics08,ES81} to derive
the set of master equations describing the time evolution of the probability distribution $\langle \rho_{\mu}\rangle$:
\begin{eqnarray}
\langle\dot{\rho}_{\mu}\rangle = \sum_{\nu}(\kappa_{\mu\nu} + \gamma_{\mu\nu}) \langle \rho_{\nu}\rangle  - \sum_{\nu}(\kappa_{\nu\mu} +
\gamma_{\nu\mu}) \langle \rho_{\mu}\rangle. \label{MasterEq}
\end{eqnarray}
where the transition matrix,
\begin{eqnarray}
\kappa_{\mu\nu} = (\kappa_{12})_{\nu\mu} + (\kappa_{23})_{\nu\mu} + (\kappa_{24})_{\nu\mu} + (\kappa_{35})_{\nu\mu} + (\kappa_{45})_{\nu\mu} +
(\kappa_{56})_{\nu\mu}, \label{KappaMuNu}
\end{eqnarray}
is represented as a sum of Marcus rates, $(\kappa_{\alpha\alpha'})_{\nu\mu}$, associated with allowed transitions between the redox states
\cite{Krish01,Marcus56,MarcusSutin85},
\begin{eqnarray}
(\kappa_{\alpha\alpha'})_{\mu\nu} = |\Delta_{\alpha\alpha'}|^2 \sqrt{\frac{\pi}{\lambda_{\alpha\alpha'}T}} [
|(a_{\alpha}^{+}a_{\alpha'})_{\mu\nu}|^2 + |(a_{\alpha}^{+}a_{\alpha'})_{\nu\mu}|^2 ] \exp\left[ - \frac{(\omega_{\mu\nu} -
\lambda_{\alpha\alpha'})^2}{4\lambda_{\alpha\alpha'}T} \right], \label{KappaAlpha}
\end{eqnarray}
where $\omega_{\mu\nu} = E_{\mu} - E_{\nu}$, and $\lambda_{\alpha\alpha'}$ is the reorganization energy corresponding to the electron transition
between $\alpha$ to $\alpha'$ redox sites \cite{Krish01,PREPump08,PumpKinetics08}.  The relaxation matrix $\gamma_{\mu\nu}$ describes a
contribution of transitions between the active sites and the electron and proton reservoirs,
\begin{eqnarray}
\gamma_{\mu\nu} = \gamma_{\rm S} \{ |a_{1;\mu\nu}|^2 [ 1 - f_{\rm S}(\omega_{\nu\mu})] + |a_{1;\nu\mu}|^2 f_{\rm S}(\omega_{\mu\nu})\} + \nonumber\\
\gamma_{\rm D} \{ |a_{6;\mu\nu}|^2 [ 1 - f_{\rm D}(\omega_{\nu\mu})] + |a_{6;\nu\mu}|^2 f_{\rm D}(\omega_{\mu\nu})\} + \nonumber\\
\Gamma_{\rm N} \{( |b_{7;\mu\nu}|^2 + |b_{8;\mu\nu}|^2 ) [ 1 - F_{\rm N}(\omega_{\nu\mu}) ] + ( |b_{7;\nu\mu}|^2 + |b_{8;\nu\mu}|^2 )F_{\rm
N}(\omega_{\mu\nu})\}
+ \nonumber\\
\Gamma_{\rm P} \{( |b_{7;\mu\nu}|^2 + |b_{8;\mu\nu}|^2 ) [ 1 - F_{\rm P}(\omega_{\nu\mu}) ] + ( |b_{7;\nu\mu}|^2 + |b_{8;\nu\mu}|^2 )F_{\rm
P}(\omega_{\mu\nu})\}. \label{gammaF}
\end{eqnarray}

\subsection{Coulomb energy and redox potential of the shuttle}
The electrostatic coupling between electrons and protons travelling together on the menaquinol molecular shuttle is of prime
importance for the electron-to-proton energy conversion.  For the sake of simplicity and without loss of generality, we describe all
electrostatic interactions involved in Eq.~(\ref{H0}) by a single electrostatic energy $u_0$: $u_{37}=u_{38}=u_{47}=u_{48}= u_0,$
and $u_{34}=u_{78}=u_0.$ It should be noted that the present model tolerates a significant spread (at least 20\% and sometimes
larger) of the electrostatic parameters. The energy scale $u_0$ is related to the redox potential $E_m$ of the MQ/MQH$_2$ couple,
which is about $- 80$ meV \cite{Blasco01}. To find this relation we model a process of redox titration of a molecule, which has one
electron and one proton-binding sites characterized by the energy levels $\varepsilon_e$ and $\varepsilon_p$, respectively.

The electron-binding site is connected to the reservoir of electrons with an electrochemical potential $\mu_e$, whereas the
protonable site is coupled to the proton reservoir with an electrochemical potential $\mu_p$. The energy of the electron-proton
Coulomb attraction is determined by the parameter $u_0$. The goal here is to determine a relation between the electron potential
$\mu_e$ and the energy scales $\varepsilon_e$ and $u_0$ when the electron-binding site is half-populated.  According to the redox
titration procedure \cite{Euro08} this value of  the ``ambient" potential $(\mu_e)_{1/2}$ determines the redox potential of the
molecule $E_m$ in the presence of electron-proton electrostatic coupling, $E_m = -\  \mu_{e,1/2}$.   As in the case of
quinone/quinol molecule the protonable site should be populated if and only if the electron-binding site is fully occupied. This
occurs at the condition: $$\varepsilon_p > \mu_p > \varepsilon_p - u_0.$$ Thus, the average electron, $\langle n_e \rangle$, and
proton, $\langle n_p \rangle$, populations of the molecule are expressed in terms of the Fermi distribution function
$f(\varepsilon)$ of the electron reservoir:
\begin{equation}
\langle n_e \rangle = \langle n_p \rangle = \frac{f(\varepsilon_e)}{1 + f(\varepsilon_e) - f(\varepsilon_e - u_0)}.
\end{equation}
The molecule is half-populated with an electron, $\langle n_e \rangle = 1/2,$ and with a proton, $ \langle n_p \rangle = 1/2$, when
\begin{equation}
\mu_{e,1/2} = -\ E_m = \varepsilon_e - \frac{u_0}{2}. \label{muU0}
\end{equation}
Calculations for a molecule having two electron sites (with energies $\varepsilon_3=\varepsilon_4 = \varepsilon_e$) and two proton-binding sites
 (with the energy levels $\varepsilon_7 = \varepsilon_8 = \varepsilon_p$) also show the validity of the relation Eq.~(\ref{muU0}) for the case of a
 single electrostatic parameter $u_0$.

\subsection{Proton-motive force}
The difference of proton electrochemical potentials, $\Delta\mu = \mu_{\rm P} - \mu_{\rm N},$ defines the transmembrane
proton-motive force, $\Delta \mu$, consisting of a voltage gradient $V$ and a contribution of the concentration difference, $\Delta
pH$, between the sides of the membrane \cite{Alberts02,Skulachev88,Nicholls92}:
\begin{equation}
\Delta \mu = V - 2.3 \,(RT/F)\times\Delta pH. \label{DMu}
\end{equation}
We introduce here the gas constant $R$ and the Faraday constant $F$. The potentials $\Delta\mu$ and $V$ are measured in meV, whereas
temperature $T$ is measured in Kelvins ($k_B=1$). At room temperature, $T$ = 298 K, and at the standard gradient of proton
concentrations, $\Delta pH = -1$ , the voltage part of the proton-motive force dominates over the contribution of the concentration
gradient: $\Delta \mu \simeq V + 60$ meV. For example, at $\Delta\mu = 200$ meV the voltage difference $V \sim 140$ meV is applied
across the membrane. As a consequence of this, the energies, $\varepsilon_{\alpha}$, of the redox sites located on the
\textit{Fdh-N} and \textit{Nar} enzymes
 are shifted from their original values $\varepsilon_{\alpha}^{(0)}$ ,
\begin{eqnarray}
\varepsilon_{\alpha} = \varepsilon_{\alpha}^{(0)} + \frac{1}{2}\; (-1)^{\alpha}\; V, \label{EnergyV}
\end{eqnarray}
where $(\alpha = 1,2,5,6)$. We assume here that  the voltage drops linearly across the membrane  \cite{Kim07}, so that the positions
of the energy levels of the electron and proton-binding sites on the shuttle are linear functions of the shuttle coordinate $x$:
\begin{eqnarray}
\varepsilon_3 = \varepsilon_4 = \varepsilon_{e}^{(0)} - \frac{x}{2 x_0}\; V, \nonumber\\
\varepsilon_7 = \varepsilon_8 = \varepsilon_{p}^{(0)} + \frac{x}{2 x_0} \;V, \label{EnergySV}
\end{eqnarray}
Here, $\varepsilon_{e}^{(0)}$ and $\varepsilon_{p}^{(0)}$ are the original values of the electron and proton energies of the
shuttle.

\subsection{Langevin equation}
Within the present model, the Brownian motion of the molecular shuttle \cite{AnnPhys05,Fox98} along a line, which connects the site 2 ($x=-x_0$)
and the site 5 ($x=x_0$), is governed by the one-dimensional overdamped Langevin equation
\begin{equation} \zeta \dot{x} = -\, \frac{dU_{c}(x)}{dx} - \langle(n_3 + n_4 - n_7 - n_8)^2\rangle \frac{dU_{s}(x)}{dx} + \xi, \label{Langevin}
\end{equation}
where $\zeta$ is the drag coefficient of the shuttle in the lipid membrane. The zero-mean valued, $\langle \xi\rangle = 0,$
fluctuation force $\xi$ has Gaussian statistics with the correlation function: $\langle \xi(t)\xi(t')\rangle~=~2 \zeta T \delta
(t-t'),$ proportional to the temperature $T$ of the environment. The diffusion coefficient $D$ of the shuttle is determined by the
Einstein relation: $D = T/\zeta.$ The potential $U_c(x)$,
\begin{equation}
U_c(x) = U_{c0} \left\{ 1  - \left[\exp\left( \frac{x - x_c}{l_c} \right) + 1 \right]^{-1} +   \left[ \exp\left( \frac{x + x_c}{l_c} \right) + 1
\right]^{-1} \right\}, \label{Uc}
\end{equation}
is responsible for the spatial confinement of the menaquinone/menaquinol molecule inside the plasma membrane with the barrier height
$U_{c0}$, the width $2x_c\ (x_c \geq x_0) $ and the steepness $l_c$. We also include in Eq.~(\ref{Langevin}) the potential $U_s(x)$
in Eq.~(\ref{Us}) hampering the Brownian motion of the charged shuttle across the lipid membrane.

\section{Parametrization of the model}

\subsection{Electron transport chain}
Within our model the electron transport chain begins with the source reservoir S characterized by the chemical potential $\mu_{\rm
S}$, which is related (with an opposite sign) to the redox energy of formate, $\mu_{\rm S} = 420$ meV \cite{Jormakka02}. The redox
potentials of hemes $b_P$ (site 1) and $b_C$ (site 2) located in formate-dehydrogenase (Fdh-N) are not known. We choose the
following values: $\varepsilon_{1}^{(0)} = 445$ meV and $\varepsilon_{2}^{(0)} = 260$ meV,  for the intrinsic energies of sites 1
and 2. Notice that with the transmembrane voltage, $V = 140$ meV, the energy (see Eq. (\ref{EnergyV}) ) of the site 1,
$\varepsilon_{1} = 375$ meV, is below the potential  $\mu_{\rm S}$, which is a necessary condition for electron transfer from the
source reservoir S to the site 1.

The original energy of electron-binding sites on the shuttle, $\varepsilon_{e}^{(0)}$, can be related to the redox potential $E_m$ of the
quinone/semiquinone (MQ$^{-}$/MQ) couple. It is known \cite{Prince83,Osyczka05} that the redox energy of the quinone/semiquinone couple is much
lower than the potential of the quinone/quinol couple. For example, the potential $E_m$ for the ubiquinone/ubiquinone (UQ/UQH$_2$) couple is
about +~60~mV, and the $E_m$ for UQ$^{-}$/UQ couple in aqueous solution is of order of $-160$ mV \cite{Nicholls92}. For the redox energy of the
MQ$^{-}$/MQ couple, we choose a value $E_m = - 215$ meV, which is below the known redox energy, $E_m = - 80$ meV, of the MQ/MQH$_2$ couple. This
means that the energy level of the electron-binding sites is placed at $\varepsilon_{e}^{(0)}$ = 215 meV. With Eq.~(\ref{muU0}) we obtain a
reasonable estimation for the charging energy of the shuttle: $$u_0 = 2\ (\varepsilon_{e}^{(0)} - \mu_{e,1/2}) = 270\,\, {\rm meV},$$ at
$\mu_{e,1/2} = - E_m$(MQ/MQH$_2$) = 80 meV. This value of the charging energy $u_0$ roughly corresponds to the electrostatic interaction of two
charges located on the opposite sides of the menaquinone molecule \cite{Maklashina06} at a distance $\sim$0.6 nm, provided that the dielectric
constant $\epsilon \sim 9$.

We note that at the voltage difference, $V = 140$ meV, the energy level of the site 2, $\varepsilon_{2}=330$ meV, is higher than the
level, $\varepsilon_{e}^{(0)} + V/2 = 285$ meV, of an electron on the shuttle located at the N-side. Because of this, electrons can
be transferred from site~2 to the menaquinone, followed by the proton uptake from the N-side of the membrane.

The unloading of the fully populated shuttle occurs at the P-side provided the energy of the electrons on the shuttle,
$\varepsilon_{e}^{(0)} - u_0 - V/2 = - 125$ meV, exceeds the energy, $\varepsilon_{5}$, of the site 5. Here, for $V= 140$~meV, we
choose sufficiently low values, $\varepsilon_{5} = - 170$ meV and $\varepsilon_{6} = - 215$ meV, for energy levels of the redox
sites 5 and 6 belonging to the second half of the redox loop, whereas the original values are $\varepsilon_{5}^{(0)} = - 100$ meV
and $\varepsilon_{6}^{(0)} = - 285$ meV. The corresponding redox potentials of these sites differ from the measured redox levels
\cite{Blasco01} of heme $b_L:\ E_m \sim 20$~mV (site 5) and heme $b_H:\ E_m \sim 120$~mV (site 6). It is known, however, that the
redox potentials obtained as a result of equilibrium redox titrations are not always applicable for a description of the electron
transfer in enzymes, in particular because of cooperativity between the redox centers \cite{Blasco01}. This cooperativity can be
 induced, e.g., by electrostatic couplings between the redox sites 1 and 2: $u_{12} = 20$ meV, and between the sites 5 and 6:
$u_{56} = 20$ meV. In the present model the electron transport chain terminates at the drain reservoir characterized (at
$V=140$~meV) by the energy scale $\mu_{\rm D} = -260$ meV, which exceeds the energy, $-E_m = - 420$ meV, of electrons at the site of
nitrate-to-nitrite reduction \cite{Blasco01}.

\subsection{Proton pathway}
Protons are loaded on the shuttle at the N-side ($x\sim - x_0$) provided that the shuttle is populated at least with one electron.
This condition can be met at $\varepsilon_{p}^{(0)} = u_0/2$ when the energy, $u_0/2 - V/2 = 65$~meV,  of a proton on the shuttle
located at $x = - x_0$, is higher than the potential $\mu_N$, whereas the proton energy level, $ - u_0/2 - V/2 = -205$~meV, of the
shuttle, populated with electrons, is below $\mu_N$. We take into account electron-electron and proton-proton Coulomb repulsions on
the shuttle and assume that $V = 140$~meV, so that the total transmembrane proton-motive force, $\Delta \mu = \mu_{\rm P} - \mu_{\rm
N}$, is about 200 meV \cite{Simon08} with $\mu_{\rm N} = -100$~meV and $\mu_{\rm P} = +100$~meV.

Unloading of protons, which occurs at the P-side of the membrane ($x\sim  x_0$), is preceded by the electron transfer to the site 5.
Then, the proton energy goes up, to the level $\varepsilon_{p}^{(0)} + V/2 = 205$~meV, exceeding the potential $\mu_{\rm P}$. It
should be noted that the present model is robust to pronounced variations ($\Delta \varepsilon \sim 50$~ meV) of electron and proton
energy levels (see Fig.~3 later on).

\subsection{Other parameters}
It is known \cite{Pilet04} that electrons can be transferred between the redox centers in a nanosecond range. The proton transfer
mediated by the hydrogen-bonded chains can occur in nanoseconds as well \cite{Nagle78,Zundel95}. In view of these findings, we
choose the following parameters controlling electron and proton transitions between the reservoirs and the active sites:
$\gamma_{\rm S} = \gamma_{\rm D} = 0.5/$ns, $\Gamma_{\rm N} = \Gamma_{\rm P} = 0.05/$ns. We assume that all allowed electron
transitions between the redox sites are determined by the same energy scale $\Delta_{\alpha\alpha'} = 8$ $\mu$eV. For the transition
lengths $l_e$ and $l_p$ involved in Eqs.~(\ref{GammaX}),\ (\ref{DeltaX}) we have the values $l_e = 0.25$~nm, $l_p = 0.25$~nm.

The reaction of the environment is described by the set of reorganization energies $\lambda_{\alpha\alpha'}$
\cite{Krish01,PREPump08,PumpKinetics08}, which are also assumed to be the same for every pair $\alpha,\alpha':$
$\lambda_{\alpha\alpha'} = \lambda = 100$~meV. A similar value of the reorganization energy has been observed in cytochrome c
oxidase \cite{Jas05}.

The Brownian motion of the shuttle is characterized by the diffusion and drag coefficients $D$ and $\zeta$. For the diffusion
coefficient we take the value $D \sim 3\cdot 10^{-12}$~m$^2$/s, measured in Ref.~\cite{Chazotte91,Marchal98} for ubiquinone
($T$~=~298 K). The drag coefficient $\zeta$ can be found from the Einstein relation, $\zeta = T/D = 1.37$~nN$\cdot$s/m. The
potential barrier $U_s(x)$ in Eq.~(\ref{Us}), which impedes the diffusion of the charged shuttle, is characterized by the energy
penalty, $U_{s0} = 770$~meV, steepness $l_s = 0.05$ nm, and half-width $x_s = 1.7$ nm. For the potential $U_c(x)$ in Eq.~(\ref{Uc}),
keeping the shuttle inside the membrane, we choose the height $U_{c0} = 500$~meV, steepness $l_c = 0.1$ nm, and half-width $x_c =
2.7$ nm. The redox sites are located at $x_0 = \pm\, 2$~nm. On average, the shuttle travels a distance $2x_0$ between sites 2 and 5
in a time $\Delta t = (2x_0)^2/(2D) \sim 2.7~\mu$s, which is much longer than the time-scales for electron and proton transitions to
and from the shuttle.

\section{Results}

To quantitatively describe the kinetics of electron and proton transfers across the membrane, we numerically solve the system of
master equations (\ref{MasterEq}) together with the Langevin equation (\ref{Langevin}) for a parameter regime, which provides a
robust and efficient proton-motive force generation, and also roughly corresponds to the menaquinone/menaquinol molecule randomly
moving inside the bacterial plasma membrane. It should be noted that the present model allows significant variations ($\sim$20\% and
sometimes higher) of the parameter values.

In Fig.~2 we present the time evolution of the electron and proton translocation process at $T$ = 298~K, $\Delta\mu = 200$~meV and
$V = 140$~meV. The shuttle starts its motion at $x=x_0$ (Fig.~2a) and after that diffuses between the membrane borders (shown by two
dashed red lines at $x = \pm \, 2$ nm). The total electron population, $n_e = \langle n_3 \rangle + \langle n_4 \rangle $
(continuous blue line), and the total proton population, $n_p = \langle n_7 \rangle + \langle n_8 \rangle$ (dashed green line), of
the shuttle is shown in Fig.~2b. The electron sites 3 and 4 are populated and depopulated in concert: $\langle n_{3}\rangle =
\langle n_{4} \rangle = n_e/2.$  The same relation takes place for the proton sites 7 and 8: $\langle n_7 \rangle = \langle n_8
\rangle = n_p/2.$ The populations are averaged over the states of electron and proton reservoirs as well as over the state of the
environment. No averaging over fluctuations of the random force $\xi(t)$ in Eq.~(\ref{Langevin}) has been performed in Fig.~2.

The total number of protons, $N_{\rm P}$ (dashed green line), transferred by the shuttle from the N- to the P-side of the membrane,
and the total number of electrons, $N_{\rm D}$ (continuous blue line), translocated from the redox site 2 to the site 5 and,
finally, to the electron drain D, are shown in Fig.~2c. At the beginning of the process ($t\sim 0$, $x \sim - x_0$) the shuttle is
rapidly populated with two electrons ($n_e =2$) and with two protons ($n_p=2$) taken from the N-side of the membrane ($\mu_{\rm N} =
-\,100$ meV). The fully loaded shuttle diffuses and eventually reaches (at $t \sim 2\,\mu$s) the opposite side where the electrons
are transferred to the redox site 5 ($N_{\rm D} = 2$), and two protons ($N_{\rm P} = 2$) are translocated energetically uphill, to
the P-side of the membrane ($\mu_{\rm P} = \,100$ meV). Accumulation of protons on the positive side of the membrane results in a
generation of the proton-motive force. The empty and neutral quinone molecule diffuses back, to the N-side of the membrane
(Fig.~2a), and the process starts again. Notice that, as a consequence of the stochastic nature of the process, the proton
population $n_p$ can be a little bit smaller than the electron population $n_e$ of the shuttle (see Fig.~2b). The resulting tiny
charge makes more difficult for the shuttle to cross the potential barrier $U_s(x)$ in Eq.~(\ref{Us}).

It is evident from Fig.~3 and Fig.~4 that the physical mechanism of proton-motive force generation described above tolerates significant
variations of system parameters such as the transmembrane voltage $V$ and temperature $T$. In Fig.~3 we show the number of protons, $N_{\rm P}$,
translocated across the membrane and the number of electrons, $N_{\rm D}$, transferred from the site 2 to the site 5 as functions of the
transmembrane voltage $V$ at $T = 298$~K.  Each point in Figs.~3 and 4 is a result of averaging over 10 realizations. Every realization has a
duration of 100~$\mu$s. We calculate the standard deviations for the number $N_{\rm P}$ of transferred protons, $\sigma_{\rm P} = \sqrt{\langle
N_{\rm P}^2\rangle - \langle N_{\rm P} \rangle^2}$, and show these deviations as the error bars in Figs.~3 and 4. The uncertainty $\sigma_{\rm
D}$ in the number $N_{\rm D}$ of translocated electrons is close to the value of $\sigma_{\rm P}$. We choose here a symmetric configuration of
the proton electrochemical potentials,
\begin{equation}
\mu_{\rm P} = -\, \mu_{\rm N} = \frac{1}{2}\  \left( V + 60 \times \frac{T}{298} \right), \label{muVT}
\end{equation}
where the potentials $\mu_{\rm N},\mu_{\rm P}, $ and the voltage $V$ are measured in meV, and the temperature $T$ is measured in
Kelvins.

It follows from Fig.~3 that this redox loop is able to translocate more than 240 protons in one millisecond against the
transmembrane voltage $V \sim 200$ meV, which corresponds to the proton-motive force $\Delta \mu \sim 260$~meV. In this case (when
$N_{\rm P} \simeq 265, \, N_{\rm D} \simeq 270,\, \mu_{\rm P} = -\mu_{\rm N} = 130$~meV, $\mu_{\rm S} = 420$~meV, and $\mu_{\rm D} =
-260$~meV) \textit{the thermodynamic efficiency} $\eta$ of the energetically uphill proton translocation,
\begin{equation}
\eta = \frac{N_{\rm P}}{N_{\rm D} } \times \frac{\mu_{\rm P} - \mu_{\rm N}}{\mu_{\rm S} - \mu_{\rm D}}, \label{eta}
\end{equation}
reaches the value $\eta \simeq 37$\%.

We note that, despite the dielectric penalty of 770 meV for a charged shuttle, the average number of transferred electrons $N_{\rm
D}$ slightly exceeds the number of protons $N_{\rm P}$. Interestingly, both numbers, $N_{\rm P}$ and $N_{\rm D}$, have small dips at
$V = 140$~meV. With increasing the transmembrane voltage, $V \geq 280$~meV, the electron transport from the site 1 ($\varepsilon_1 =
305$) to the site 2 ($\varepsilon_2 = 400$), and from the site 5 ($\varepsilon_5 = -240$) to the site 6 ($\varepsilon_6 = -145$, all
energies in meV) become energetically unfavorable. As a result of this, the numbers of electrons, $N_{\rm D}$, and protons, $N_{\rm
P}$, translocated across the membrane drop significantly at high voltages.

The temperature dependence of the average numbers of protons, $N_{\rm P}$, and electrons, $N_{\rm D}$, conveyed by the shuttle is
presented in Fig.~4 for $V = 140$~meV. The system demonstrates stable performance with $N_{\rm P} \sim 220$ protons/ms in a window
of temperatures from 250 K up to 350 K. The initial increase of $N_{\rm P}$ and $N_{\rm D}$ with temperature is probably due to the
fact that in a warmer environment the shuttle travels more frequently between the sides of the membrane transferring more electrons
and more protons. Loading (unloading) the shuttle with protons follows its loading (unloading) with electrons. At high temperatures
menaquinone spends less time in the loading zone (at $x \sim - x_0$), and protons have less opportunity to populate the shuttle.
Therefore, the gap between the numbers of transferred protons and electrons widens with increasing temperature. This means that at
high temperatures the shuttle has more chances to carry a charge, which obstructs the shuttle's diffusion across the membrane.
Besides that, at sufficiently high temperatures electrons have not enough time to be loaded on the shuttle. A combination of these
two features results in the high-temperature decline of electron and proton flows shown in Fig.~4.

\section{Conclusions}

Using a simple kinetic model we have examined the process of proton-motive force generation across the bacterial plasma membrane. This model is
applied to the redox loop mechanism of the nitrate respiration in \textit{E.~coli}. This approach includes two redox sites in the first half of
the redox loop, two redox sites in the second half, and the Brownian shuttle diffusing between the negative (N) and positive (P) sides of the
membrane. We show that the Coulomb attraction between electrons and protons travelling on the shuttle plays an essential role in the
energetically-uphill proton translocation from the N-side to the P-side of the membrane and, thus, in the proton-motive force generation. We
have derived and numerically solved a  set of master equations, which quantitatively describes the process of loading and unloading the shuttle
with electrons and protons, along with a stochastic Langevin equation for the shuttle position. Our model is able to explain the generation of
the proton-motive force up to 300 meV in the physiologically relevant range of temperatures from 250 to 350 K with a peak thermodynamic
efficiency of about 37\%. A sequence of electron and proton transport events and main characteristics of the redox loop mechanism calculated in
the present paper can be measured in future experiments aimed on a kinetic analysis of the nitrate respiration process in bacteria.

\section{Acknowledgements} This work was supported in part by  the National Security Agency (NSA), Laboratory of Physical Science (LPS), Army
Research Office (ARO), National Science Foundation (NSF) grant No. 0726909.
 S.S. acknowledges support from the EPSRC via EP/D072581/1.

\newpage

\begin{figure}
\includegraphics[width=14.0cm, height=15.0cm ]{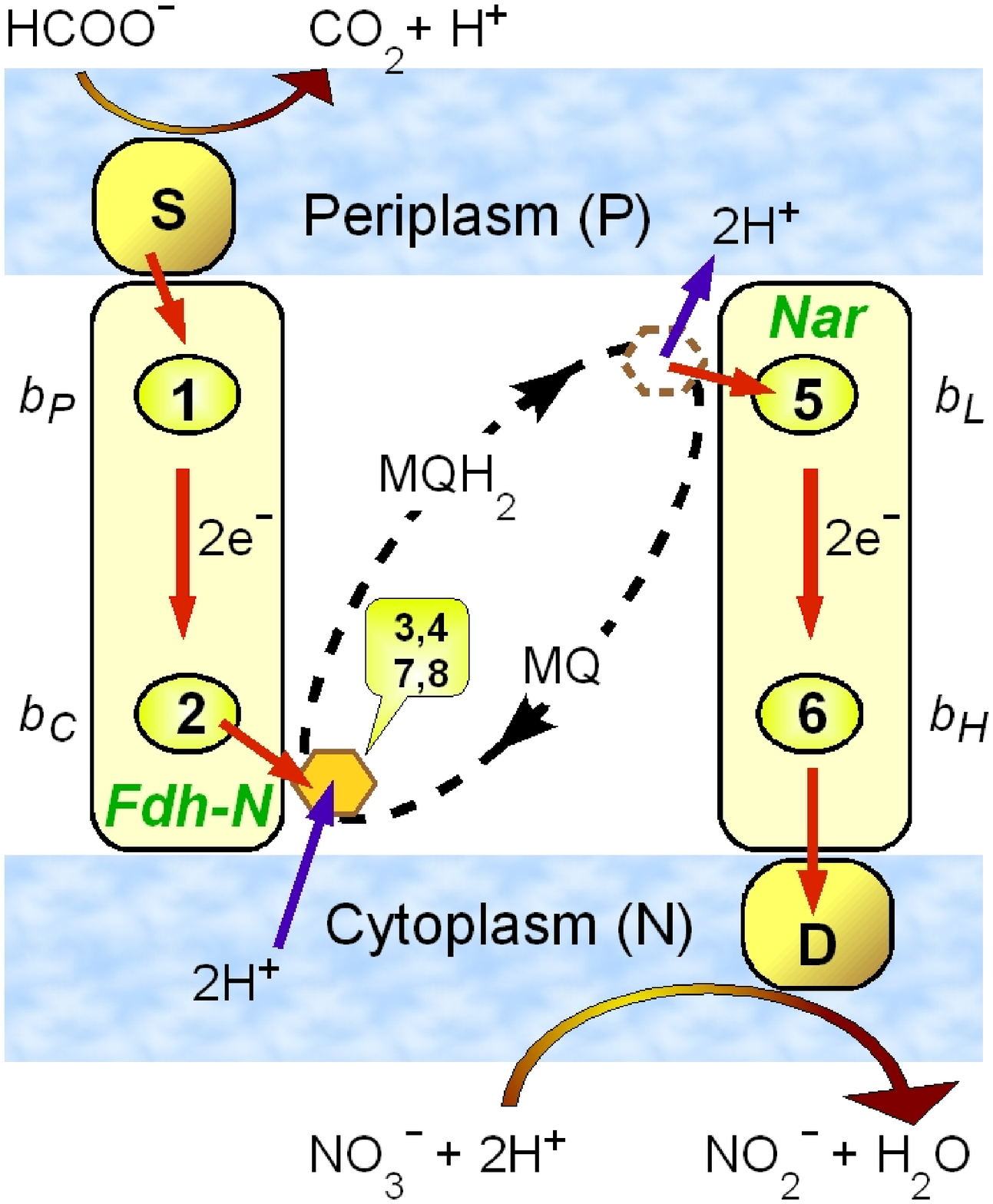}
\vspace*{-2cm} \caption{ (Color online) Schematic diagram of the redox loop. High-energy electrons are delivered from the source S
to a redox center 1 (heme $b_{P}$) located near the periplasmic (P) side of the membrane. After that, electrons are transferred
across the membrane to a redox site 2 (heme $b_{C}$) on the cytoplasmic (N) side. At the N-side two electrons reduce a molecule of
menaquinone MQ, which also takes two protons from the N-side turning into a molecule of menaquinol MQH$_2$. The menaquinone shuttle
has two electron-binding sites, 3 and 4, and two protonable sites, 7 and 8. The neutral quinol molecule MQH$_2$ diffuses freely to
the P-side of the membrane, where its electron cargo is transferred to the redox site 5 (heme $b_L$), and, via the center 6 (heme
$b_H$), to the drain D on the cytoplasmic side. The oxidation of the quinol molecule MQH$_2$ by the center 5 is accompanied by a
release of two protons to the P-side of the membrane. Formate dehydrogenase (\textit{Fdh-N}, with centers $b_P$ and $ b_C$) reduces
the quinone molecule MQ. Nitrate reductase (\textit{Nar}, with centers $b_L$ and $b_H$) oxidizes the quinol molecule MQH$_2$. Both
of these (\textit{Fdh-N} and \textit{Nar}) form the redox loop, generating a proton-motive force across the membrane.}
\end{figure}

\begin{figure}
\includegraphics[width=20.0cm, height=12cm ]{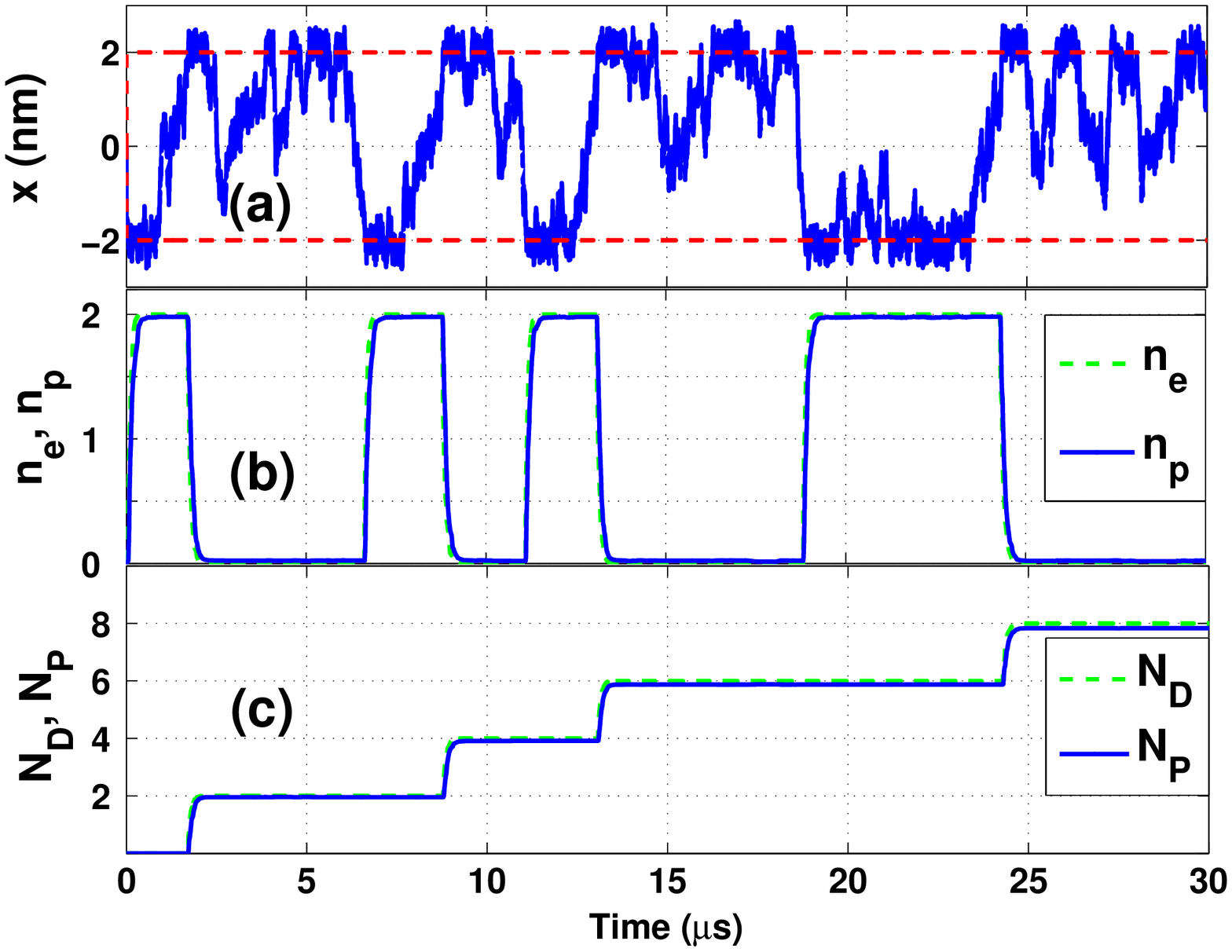}
\vspace*{-1cm} \caption{ (Color online) (a) Time dependence of the position $x$ (in nm) (blue continuous curve) of the shuttle, diffusing
between the walls of the plasma membrane located at $x = \pm x_0$ (two red dashed horizontal lines), where $x_0$ = 2 nm;
 (b) the total proton ($n_p = \langle n_7\rangle + \langle n_8\rangle$, blue continuous curve)
 and electron ($n_e = \langle n_3\rangle + \langle n_4\rangle$, green dashed curve)
populations of the shuttle versus time (in $\mu$s);
 (c) the number of transferred protons ($N_{\rm P}$, blue continuous curve) and the number of translocated electrons ($N_{\rm D}$, green dashed curve)
 versus time at $V = 140$~meV, $\Delta \mu = 200$~meV, and at $T$ = 298 K. Notice that the shuttle is loaded near the N-side
 of the membrane, at $x \approx -~x_0$, and unloaded at the P-side, at $x \approx +~x_0$. It follows from (c) that the process of shuttle
 unloading is accompanied  by a stepwise increase of the number of protons, $N_{\rm P}$, translocated to the P-side of the membrane, and the
 number of electrons, $N_{\rm D}$, transferred to the site 5 and, finally, to the drain.}
\end{figure}

\begin{figure}
\includegraphics[width=20.0cm ]{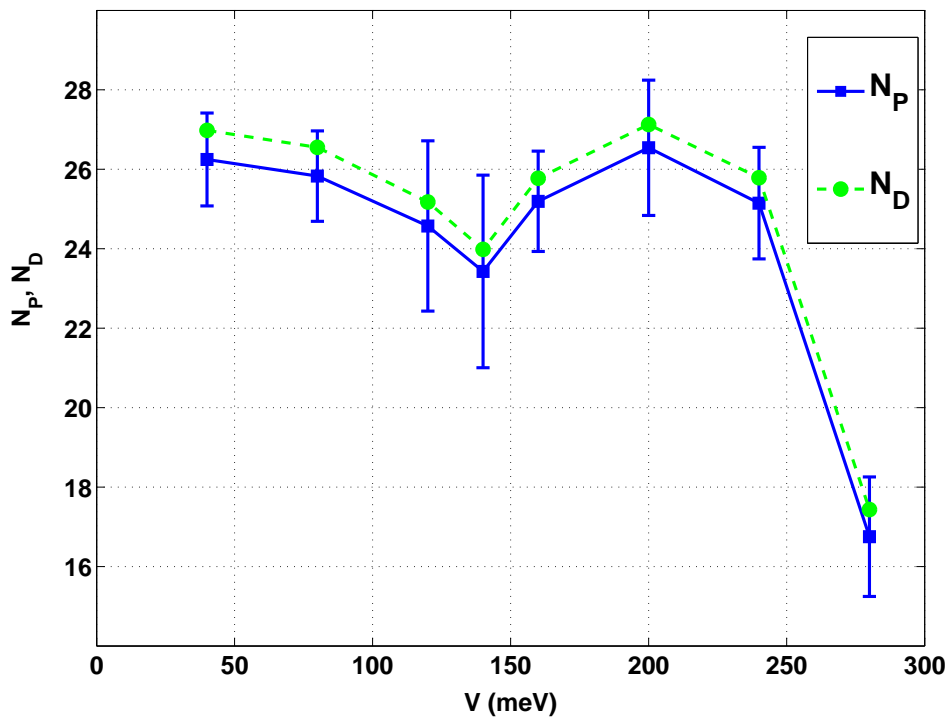}
\vspace*{0cm} \caption{ (Color online) The number of protons, $N_{\rm P}$ (blue continuous curve), translocated energetically uphill, from the
N-side to the P-side of the membrane, and the number of electrons, $N_{\rm D}$ (green dashed curve), transferred from the site 2 on the $Fdh-N$
enzyme to the site 5 belonging to the $Nar$ enzyme, as functions of the transmembrane voltage $V$ at $T$ = 298 K. In Figs.~3 and 4 the results
are averaged over 10 realizations. Each realization has a time span of 100 $\mu$s. Error bars (standard deviations) are shown for the number
$N_{\rm P}$ of translocated protons.}
\end{figure}

\begin{figure}
\includegraphics[width=20.0cm ] {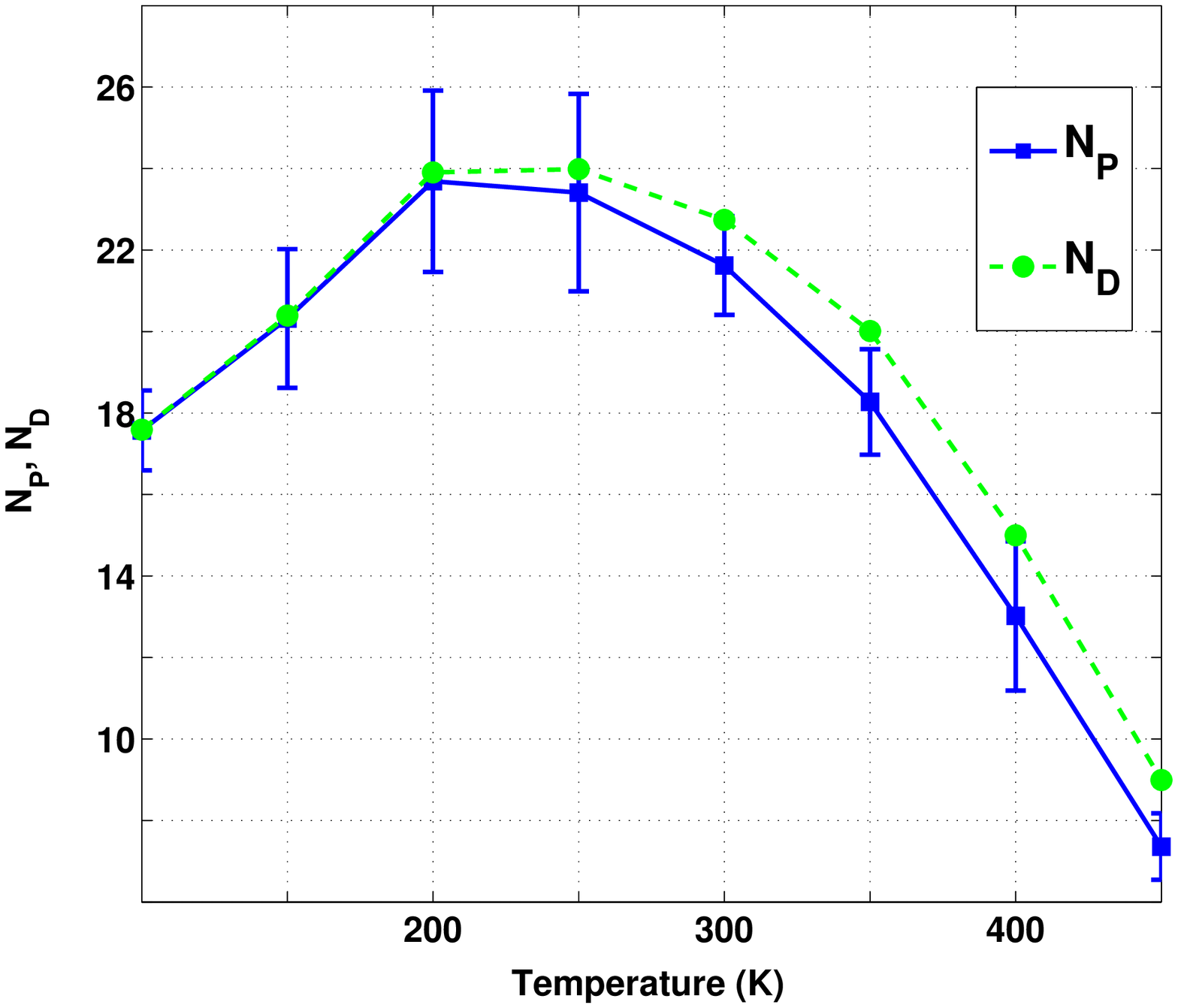}
\vspace*{0cm} \caption{ (Color online) Temperature dependence of the numbers of protons, $N_{\rm P}$ (blue continuous curve, with error bars)
and electrons, $N_{\rm D}$ (green dashed curve), transported by the shuttle at the transmembrane voltage $V = 140$~meV.}
\end{figure}

\end{document}